\documentclass[11pt,letterpaper]{article}

\usepackage{mathptmx}       
\usepackage{helvet}         
\usepackage{courier}        
\usepackage{type1cm}        
\usepackage{makeidx}         
\usepackage{graphicx}        
\usepackage{multicol}        
\usepackage[bottom]{footmisc}

\usepackage{graphicx}
\usepackage{epstopdf}
\usepackage{mathrsfs}
\usepackage{bbding}
\usepackage{subfigure}
\usepackage{float}
\usepackage{undertilde}
\usepackage{epstopdf}
\usepackage{url}
\usepackage{enumitem}
\usepackage{textcomp}
\usepackage{color}

\usepackage{setspace}
\usepackage[left=2cm, right=2cm, top=2.50cm, bottom=2.00cm]{geometry}

\begin{document}

\begin{center}
\Large{Temperature Induced Cubic-to-Tetragonal Transformations in\\ Shape Memory Alloys Using a Phase-Field Model} 
\end{center}

\begin{center}
R. Dhote$^{1,3}$, H. Gomez$^2$, R. Melnik$^3$, J. Zu$^1$

$^1$Mechanical and Industrial Engineering, University 
of Toronto, \\5 King's College Road, Toronto, ON, M5S3G8, Canada\\
$^2$Department of Applied Mathematics, University of A Coru\~{n}a \\ Campus de Elvina, s/n. 15192 A Coru\~{n}a, Spain\\
$^3$M$^2$NeT Laboratory, Wilfrid Laurier University, Waterloo, ON,  
N2L3C5, Canada
\end{center}

\abstract{
Shape memory alloys (SMAs) exhibit hysteresis behaviors upon stress and temperature induced loadings. In this contribution, we focus on numerical simulations of microstructure evolution of cubic-to-tetragonal martensitic phase transformations in SMAs in 3D settings under the dynamic loading conditions. A phase-field (PF) model has been developed to capture coupled dynamic thermo-mechanical behavior of such SMA structures and the system of governing equations have been solved numerically using the isogeometric analysis. Temperature induced reverse and forward transformations have been applied to a cubic SMA specimen, starting with evolved accommodated martensitic microstructure. We have observed that during the forward transformation, the martensitic variants nucleate abruptly. The transient microstructures are aligned along $\left[  110 \right]  $ planes, which is in accordance with the crystallographic theory and experimental results.   
}

\section{Introduction} \label{sec:Introduction}
SMAs have been widely used in commercial applications and studied in the research community for their interesting shape recovering, hysteretic properties and complex microstructure morphology \cite{Kaushik,Otsuka,Lagoudas}. Most of these studies/applications have been developed to model/utilizes static or quasi-static behaviors of SMAs. There exist a number of areas (e.g.  energy absorption and vibration damping, to just name a few)  where the dynamic behavior of SMAs is essential. Our better understanding of microstructure evolution and its effect on SMA properties, upon dynamic loading, will help in the development of better models and devices. 

In this contribution, we present a 3D model to study cubic-to-tetragonal phase transformations in SMAs. The model is developed based on a phase-field approach and the phenomenological Ginzburg-Landau theory \cite{Barsch1984,Ahluwalia2006,dhote2013PCS}. A Ginzburg-Landau free energy functional is defined in terms of two (deviatoric) strain based order parameters, whose roots define a phase in a system at a particular temperature. The austenite phase is represented by a cubic arrangement of atoms which occur at higher temperatures. The tetragonal arrangements of atoms occur at lower temperatures resulting in martensitic variants, which  are energetically equivalent.  The governing equations of the mathematical model are derived from the conservation laws of mass, momentum, and energy \cite{Melnik2002}. The developed model has highly nonlinear hysteretic behavior, bi-directional thermo-mechanical coupling and higher (fourth) order spatial differential terms \cite{dhote2013PCS}. The fourth order differential terms define a smoothly varying diffuse interface between austenite and martensite variants or between martensite variants. Traditionally, such higher order differential models have been numerically solved using a finite difference, spectral methods, etc. \cite{Ahluwalia2006}. These methods have known limitations in terms of geometric flexibility of a domain.  An isogeometric analysis (IGA) is a geometrically flexible method that can be used to study real world devices of complex shape. IGA   offers advantages in exact geometric representations, higher-order continuity, accuracy and robustness \cite{Hughes}. In  \cite{dhote2013PCS}, we first reported the use of IGA methodology to study microstructure evolution for the 3D cubic-to-tetragonal phase transformations in SMAs. In this paper, we study microstructure evolution in SMAs under temperature induced transformations.

In Section 2, we present the phase-field model describing the cubic-to-tetragonal transformations in SMAs and its numerical implementation based on the IGA. In Section 3, we study microstructure evolution in a SMA domain under reverse and forward transformations starting with accommodated twinned microstructures.  Conclusions are discussed in Section 4. 
 
\section{Mathematical Model and Numerical Implementation} \label{sec:Model}
The following three-well Ginzburg-Landau free energy functional can be used to describe cubic-to-tetragonal phase transformations in SMAs. The functional is expressed in terms of strain based order parameters and temperature, as
\begin{equation}
\mathscr{F} = \frac{a_{31}}{2} \left[ e_1^2 \right] + \frac{a_{36}}{2} \left[ e_4^2 + e_5^2 + e_6^2 \right] + 
\frac{a_{32}}{2} \tau (e_2^2+e_3^2) + \frac{a_{33}}{2} e_3 (e_3^2-3 e_2^2) + \frac{a_{34}}{2} (e_2^2 + e_3^2)^2 + 
\frac{k_g}{2} \left[ (\nabla e_2)^2 + (\nabla e_3)^2 \right],\nonumber\\
\end{equation} 
where $ a_{ij} $, $ k_g$ are the material parameters and $ \tau $ is the temperature coefficient \cite{Barsch1984,Ahluwalia2006,dhote2013PCS}. The strain $ e_1 $ represents bulk strain, $ e_2 $ and $ e_3 $ represent deviatoric strains, and $ e_4, e_5, $ and $ e_6 $ represent shear strains. The $ e_i $ strains are defined using the Cauchy-Lagrange strain tensor as  $e_{ij} = \left[\left(\partial u_{i}/\partial x_{j}\right) + \left(\partial u_{j}/\partial x_{i}\right)\right]/2$ (using the repeated index convention), where $\textbf{u}=\{u_i\}|_{i=1,2,3}$ are the displacements along \textit{x}, \textit{y}, and \textit{z} directions, respectively. The first and second terms in the functional represent bulk, and shear energy, respectively. The next three terms represent the Landau energy that define phase transformations between austenite and martensites and between martensite variants. The last term represents the gradient energy that describes non-local elastic behavior. The Landau energy has three minima having equal energies, corresponding to the three martensitic variants, below the critical temperature, one minima, corresponding to the austenite phase, above the critical temperature. The system has degenerate state near the critical temperature. 

The mathematical model is described by conservation laws of mass, momentum, and energy \cite{Melnik2002,dhote2013PCS} as
\begin{eqnarray}
&&\vec{\dot{u}} = \vec{v}, \label{eq:dummyeqn} \\
&&\rho \vec{\dot{v}} = \nabla \cdot \vec{\sigma} + \nabla \cdot \vec{\sigma}^{\prime}  + \vec{f}, \label{eq:mechanicaleqn}\\
&&\rho \dot{e} -  \vec{\sigma}^T : (\nabla \vec{v}) + \vec{\nabla} \cdot \vec{q} = g, \label{eq:thermaleqn}
\end{eqnarray}
where $ \rho $ is the mass density, $ \vec{q} $ is the Fourier heat flux vector, $ \vec{f} $, and $ g $ are external mechanical and thermal loadings. The stress tensors $ \vec{\sigma} $ and dissipation stress tensors $ \vec{\sigma}^{\prime} $ are defined as
\begin{eqnarray}
\vec{\sigma} = \frac{\partial \mathscr{F}}{\partial e_{ij}}, \qquad
\vec{\sigma}^{\prime} = \frac{\partial \mathscr{R}}{\partial \dot{e}_{ij}}.
\end{eqnarray}
The Rayleigh dissipation energy functional $ \mathscr{R} = \eta/2 \sum \dot{e}_i^2 $ is added to stabilize the microstructure quickly, where $ \eta $ is the dissipation coefficient. 

The developed model has highly nonlinear hysteretic behavior, thermo-mechanical coupling and fourth order spatial differential terms. The weak form of the governing Eqs. (\ref{eq:dummyeqn})-(\ref{eq:thermaleqn}) are obtained by multiplying them with weighting functions and transforming them by using the integration by parts. We implement the weak form of the governing equations in the isogeometric analysis for numerical solution. The semi-discrete formulation, where the space is discretized using the Galerkin finite element scheme and time is treated as continuous has been described in \cite{dhote2013PCS}. 

\section{Numerical Simulations} \label{sec:Simulations}

The simulations in this section are conducted on a cubic domain with 80 nm side. All the simulations have been performed on the Sharcnet clusters utilizing 64 processors (4 processors each in three directions) with 1 GB memory each. The decomposed domain, in each processor, is discretized with 16 quadratic $\mathscr{C}^1 $-continuous NURBS basis in each direction. The periodic boundary conditions have been used in the structural physics and insulated for the thermal physics. The material parameters are identical to those used in \cite{Ahluwalia2006}. The simulations have been carried out to study microstructure evolution under temperature induced reverse and forward phase transformations, without application of a mechanical load.

\begin{figure}[H]
	\centering
	\subfigure[M$_1$]
	{
	\includegraphics[width=0.25\textwidth]{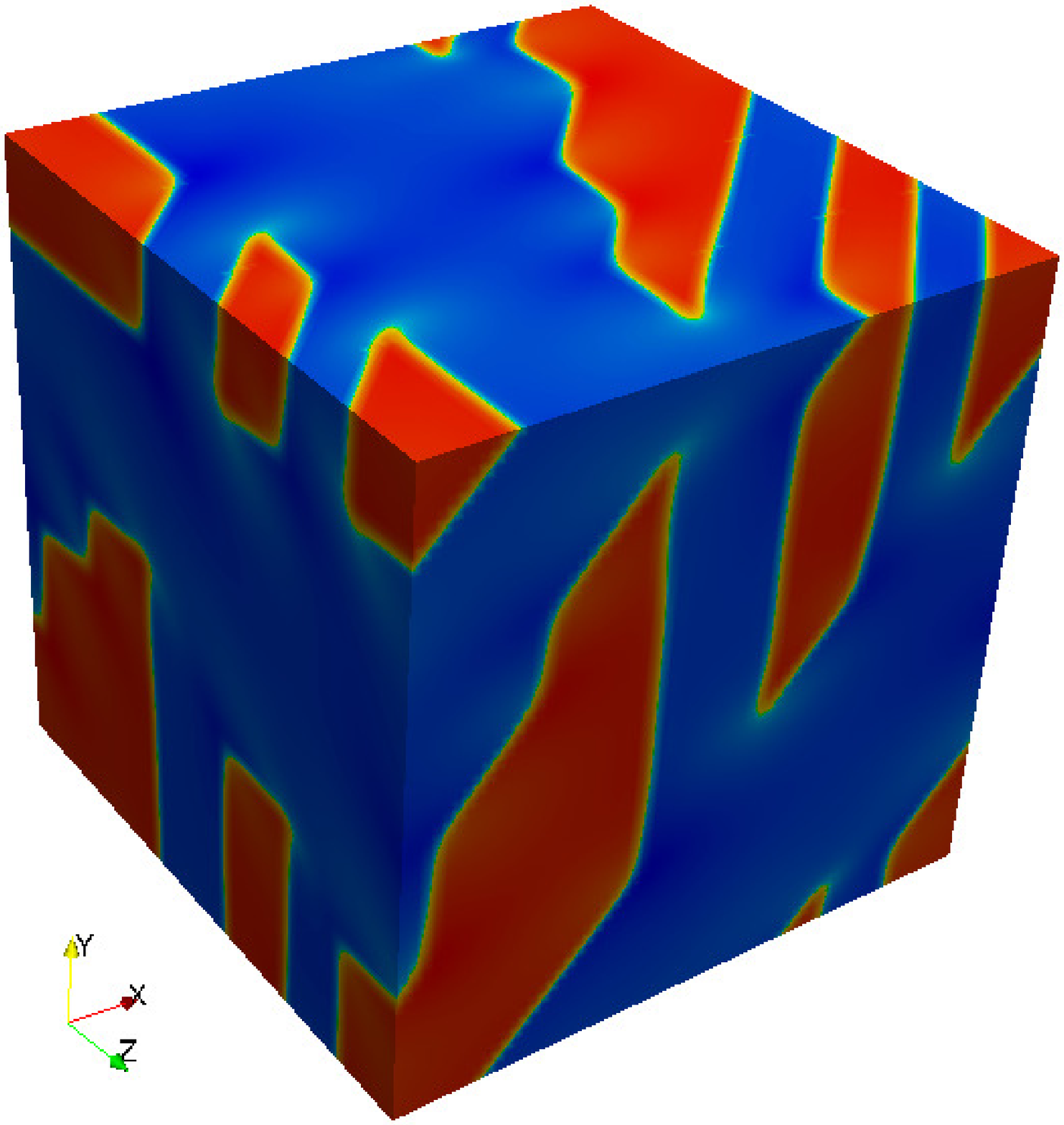} 
	   \label{fig:uxevolve}
	}
	\subfigure[M$_2$]
	{
	\includegraphics[width=0.25\textwidth]{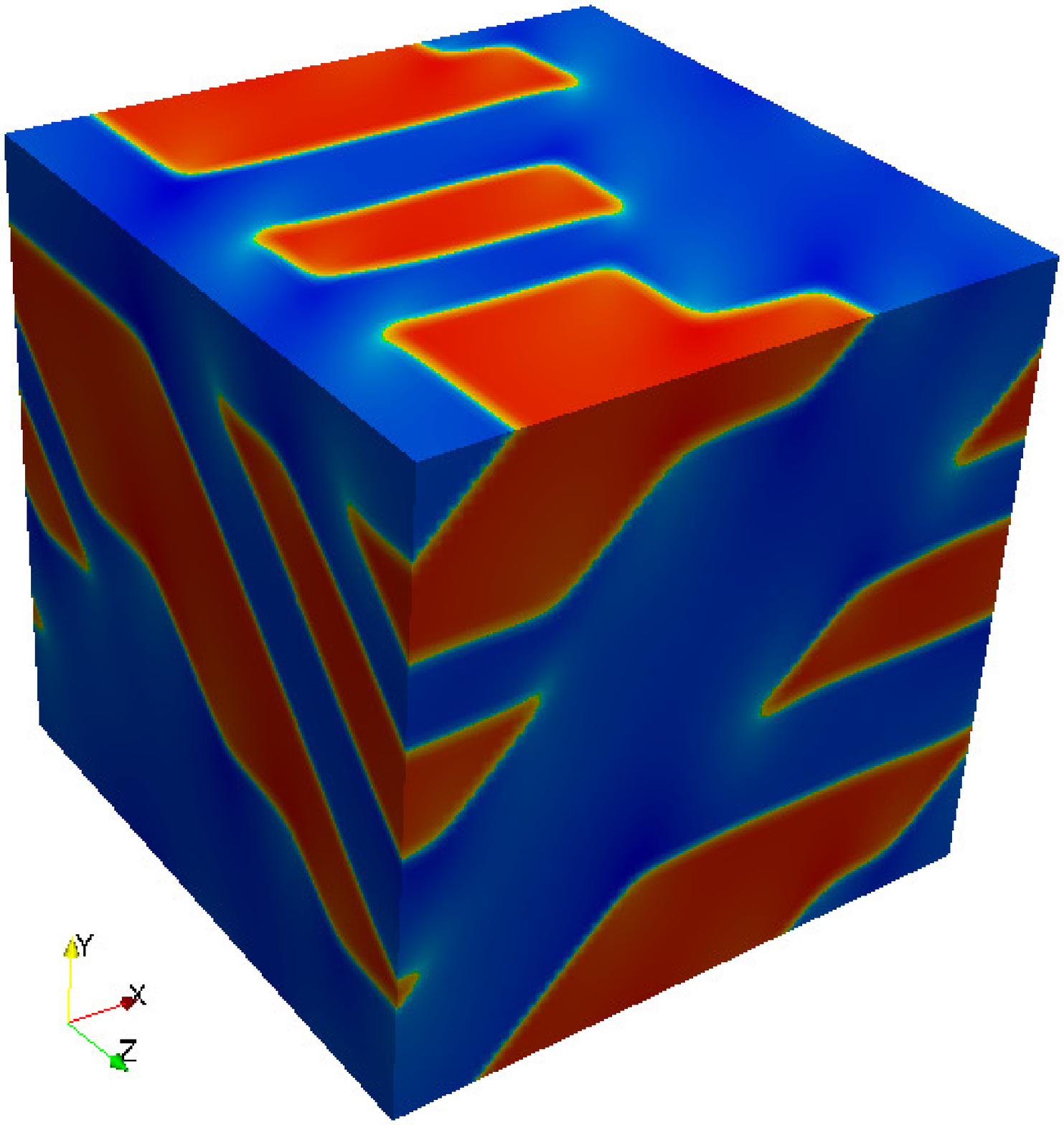} 
	   \label{fig:vyevolve}
	}	
	\subfigure[M$_3$]
	{
	\includegraphics[width=0.25\textwidth]{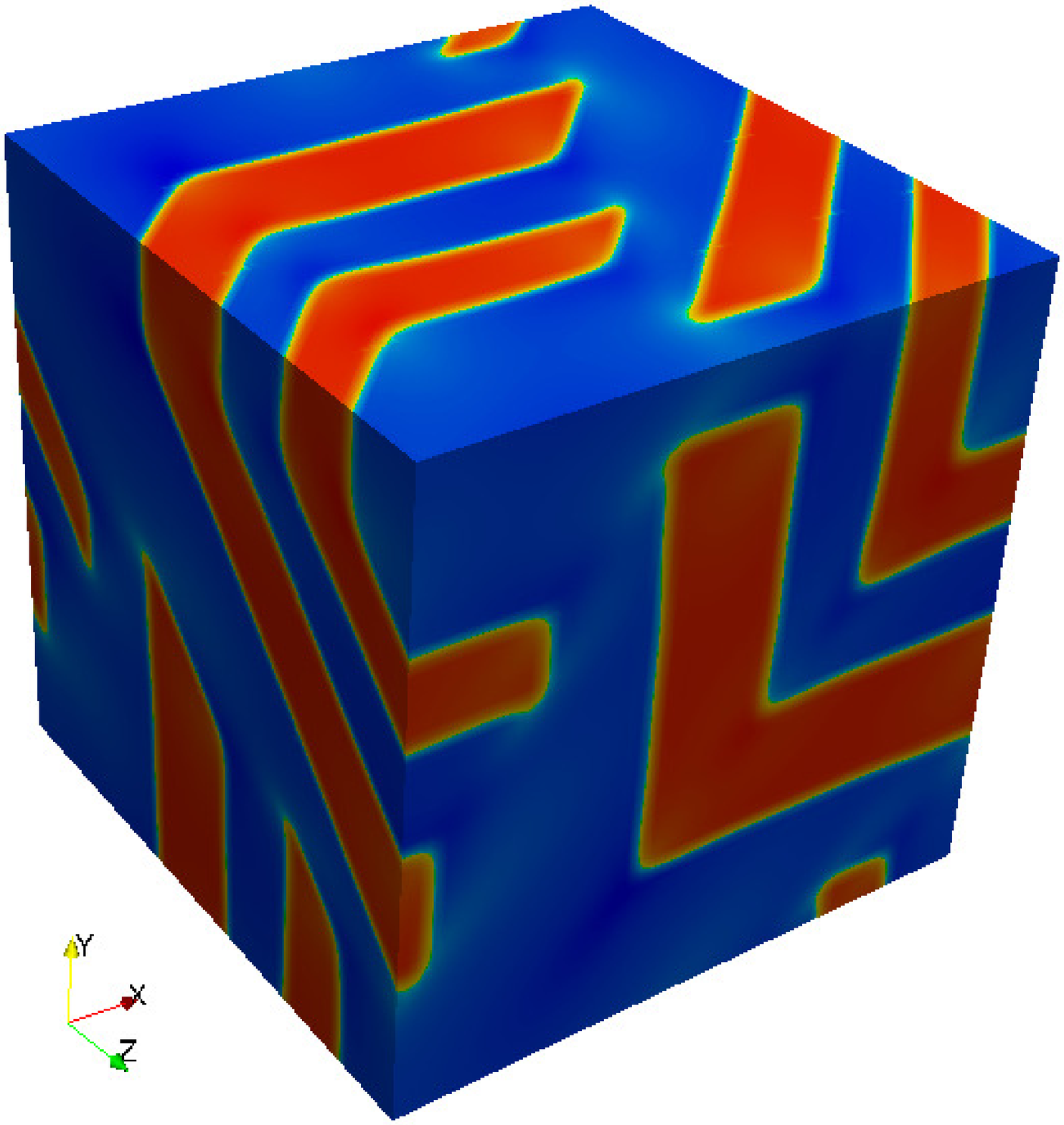} 
	   \label{fig:wzevolve}
	}
	\subfigure
	{
	\includegraphics[width=0.25\textwidth]{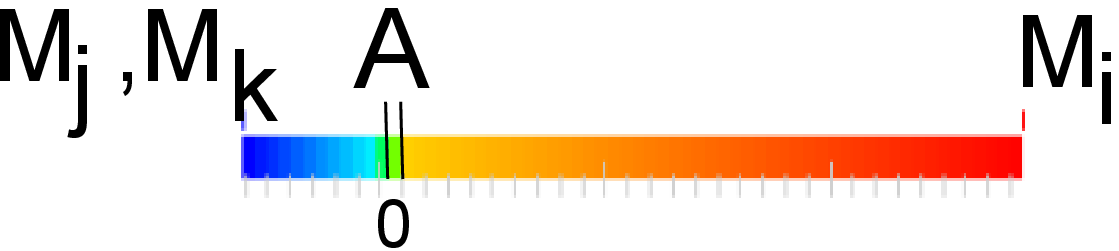} 
	}			
	\caption{(Color online) Self accommodated microstructure in a cube domain with (a) M$_1 $, (b) M$_2 $, (c) M$_3 $ martensitic variants (red color represent  M$_i $ variant, blue represent the remaining two variants M$ _j $ and M$_k $, and green color represents austenite (A) phase).}
    \label{fig:MicrostructureEvolve}    
\end{figure}

We first obtain the accommodated twinned microstructure in a domain by allowing the system to evolve, starting with initial random conditions in displacement $ \vec{u} $ and temperature coefficient $ \tau = -1.2 $. The system minimizes its energy and stabilizes into accommodated twinned martensitic variants. Fig. \ref{fig:MicrostructureEvolve} shows the three variants of martensites M$_1 $, M$_2 $, and M$_3 $ corresponding to martensite phase (tetragonal) aligned along the x, y and z, directions, respectively. The microstructures are characterized by the axial strain values (e.g. martensite M$_1 $ is represented by $ \epsilon_{11} > 0 $ i.e. tetragonal variant  elongated along the x-direction). The red color in each subplot of Fig. \ref{fig:MicrostructureEvolve} represents M$_i $ variant and blue represents the remaining two variants, as shown in the color spectrum in the figures. The competition between bulk, shear, and gradient energy results in three variants accommodated in a herringbone structure with domain walls aligned along $\left[ 110 \right] $ planes, which is in accordance with the crystallographic theory and experimental results \cite{Kaushik,Sapriel1975}.

\begin{figure}[H]
	\centering
	\subfigure
	{
	\includegraphics[width=0.25\textwidth]{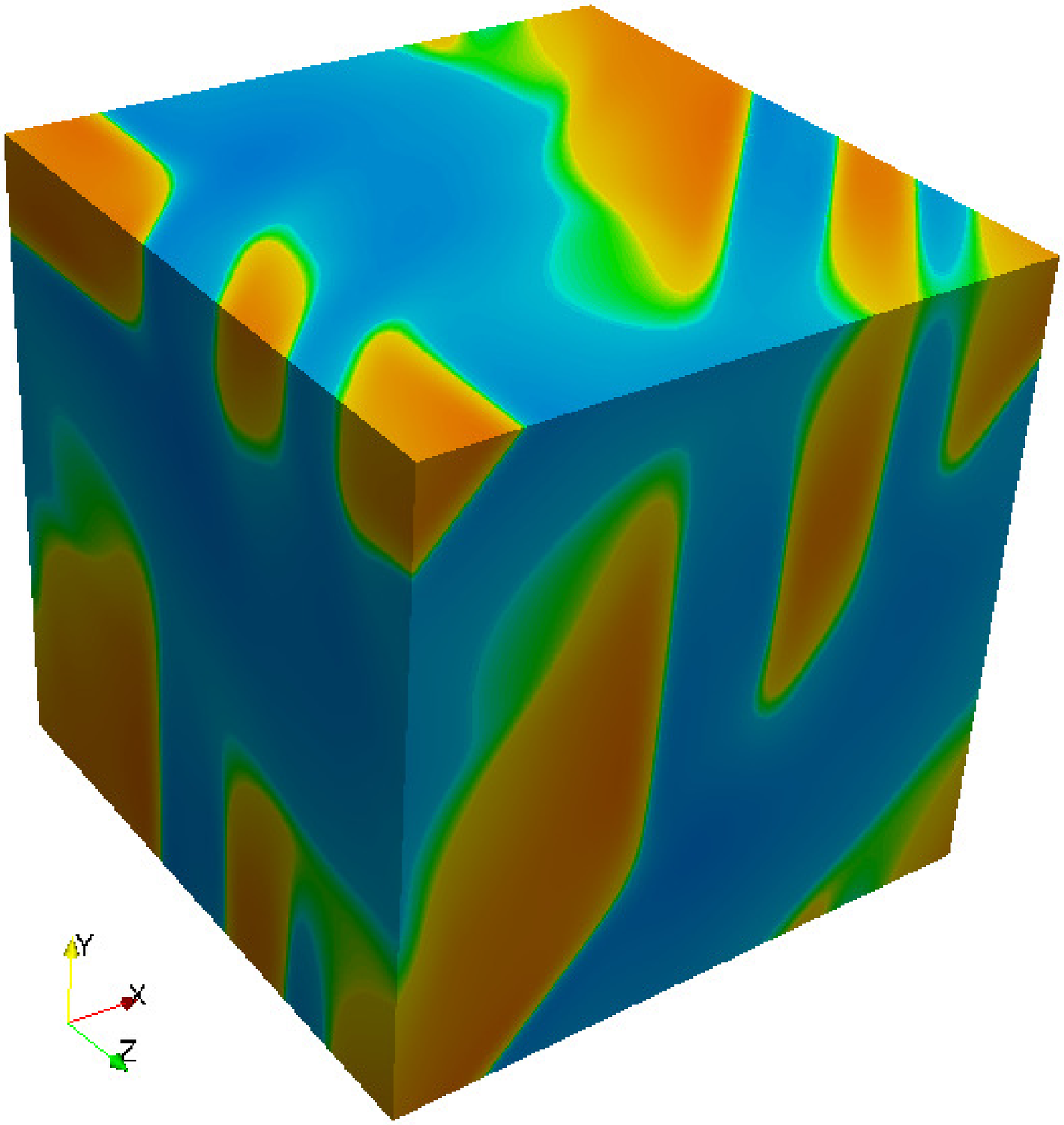} 
	   \label{fig:loadux1}
	}
	\subfigure
	{
	\includegraphics[width=0.25\textwidth]{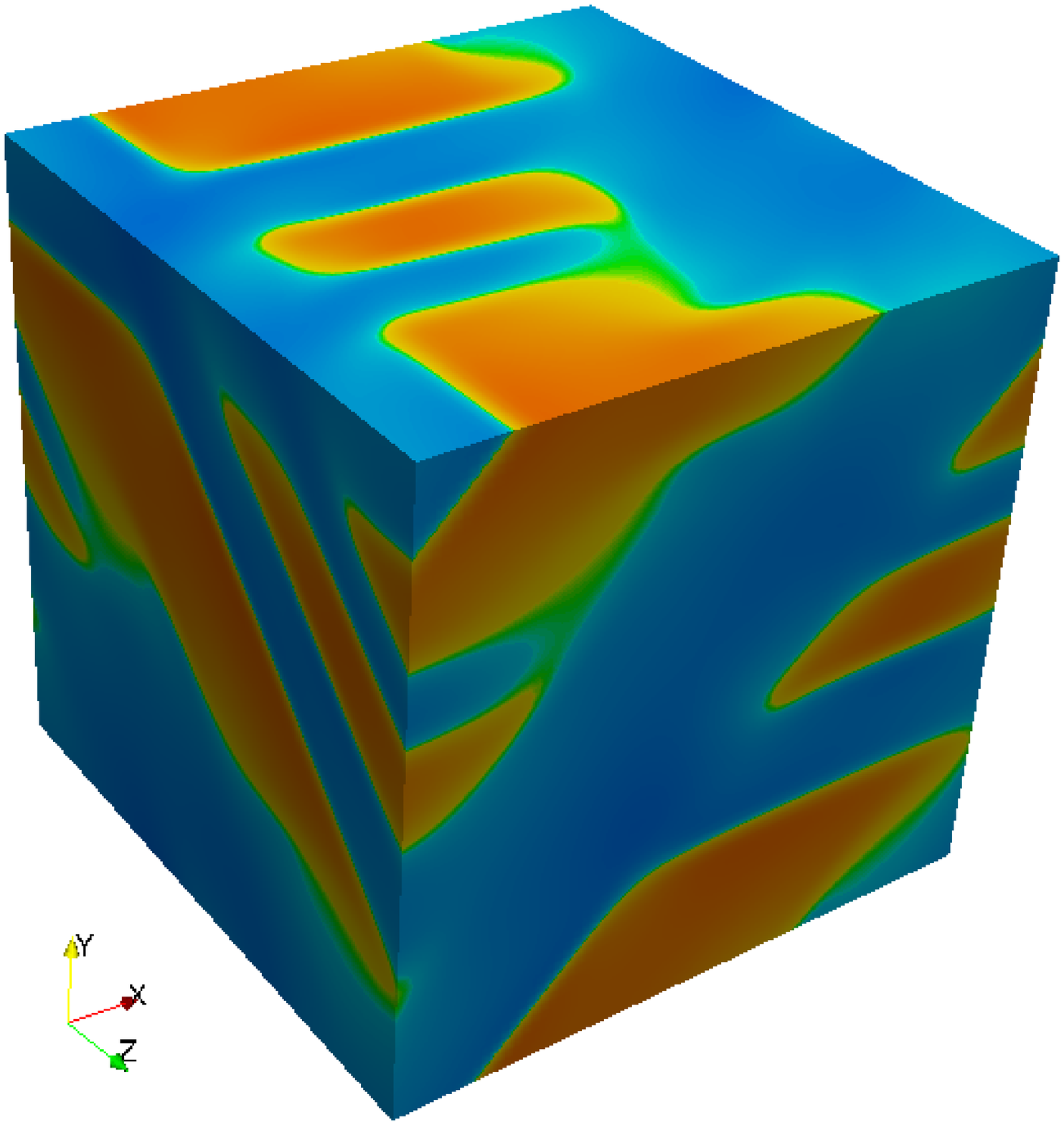} 
	   \label{fig:loadvy1}
	}	
	\subfigure
	{
	\includegraphics[width=0.25\textwidth]{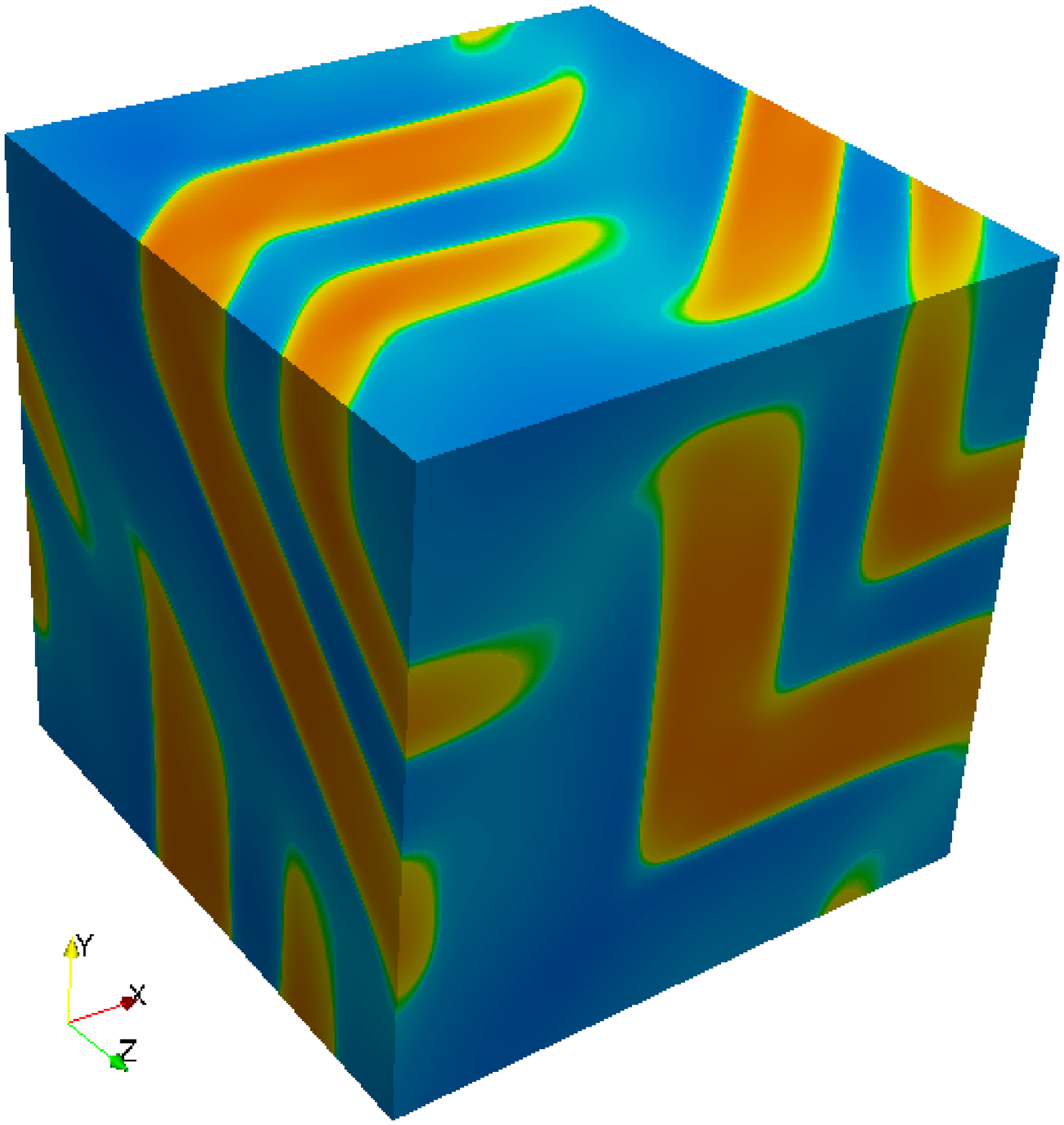} 
	   \label{fig:loadwz1}
	}	\\	
	\setcounter{subfigure}{0}
	\subfigure[M$_1$]
	{
	\includegraphics[width=0.25\textwidth]{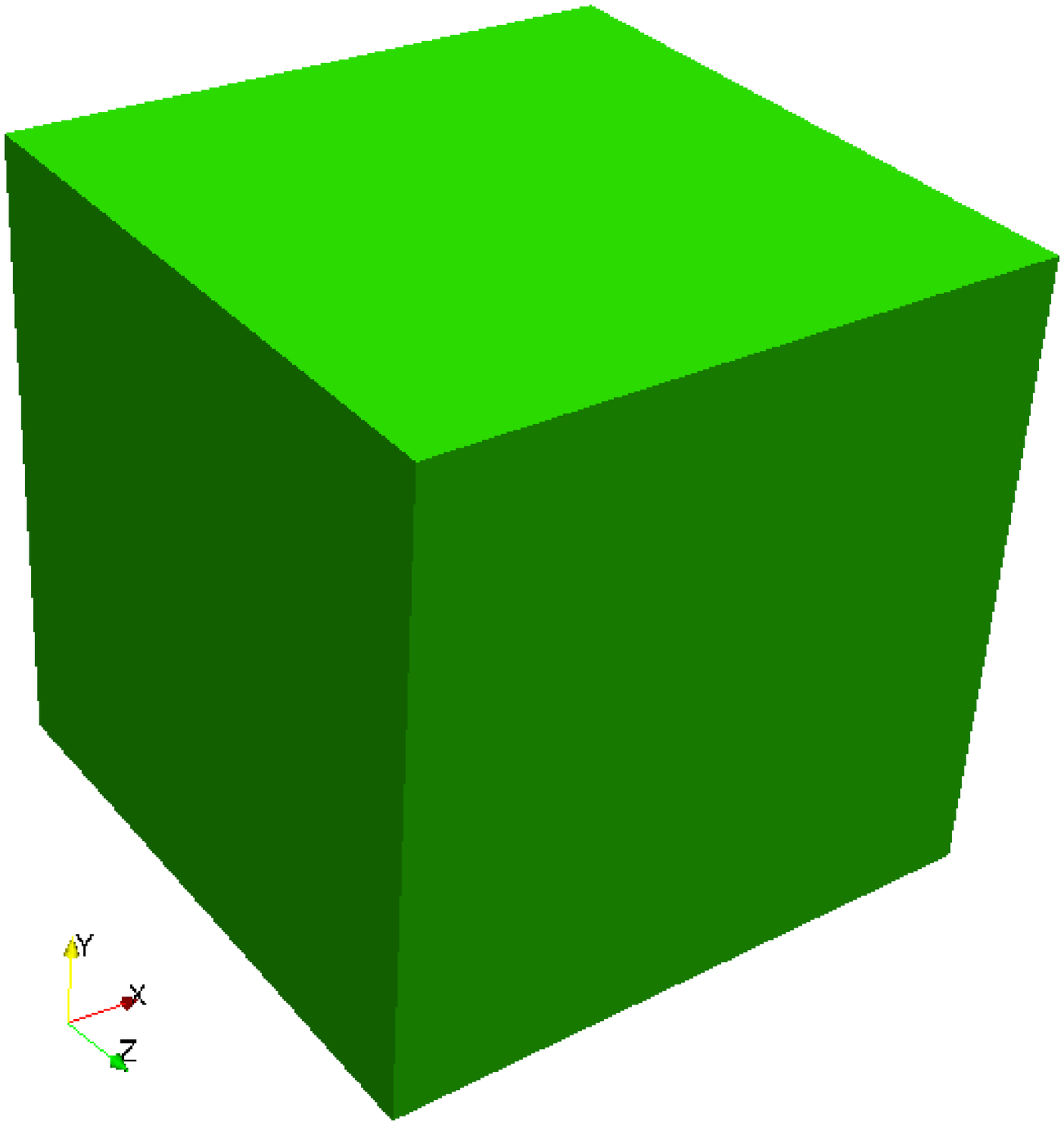} 
	   \label{fig:loadux2}
	}
	\subfigure[M$_2$]
	{
	\includegraphics[width=0.25\textwidth]{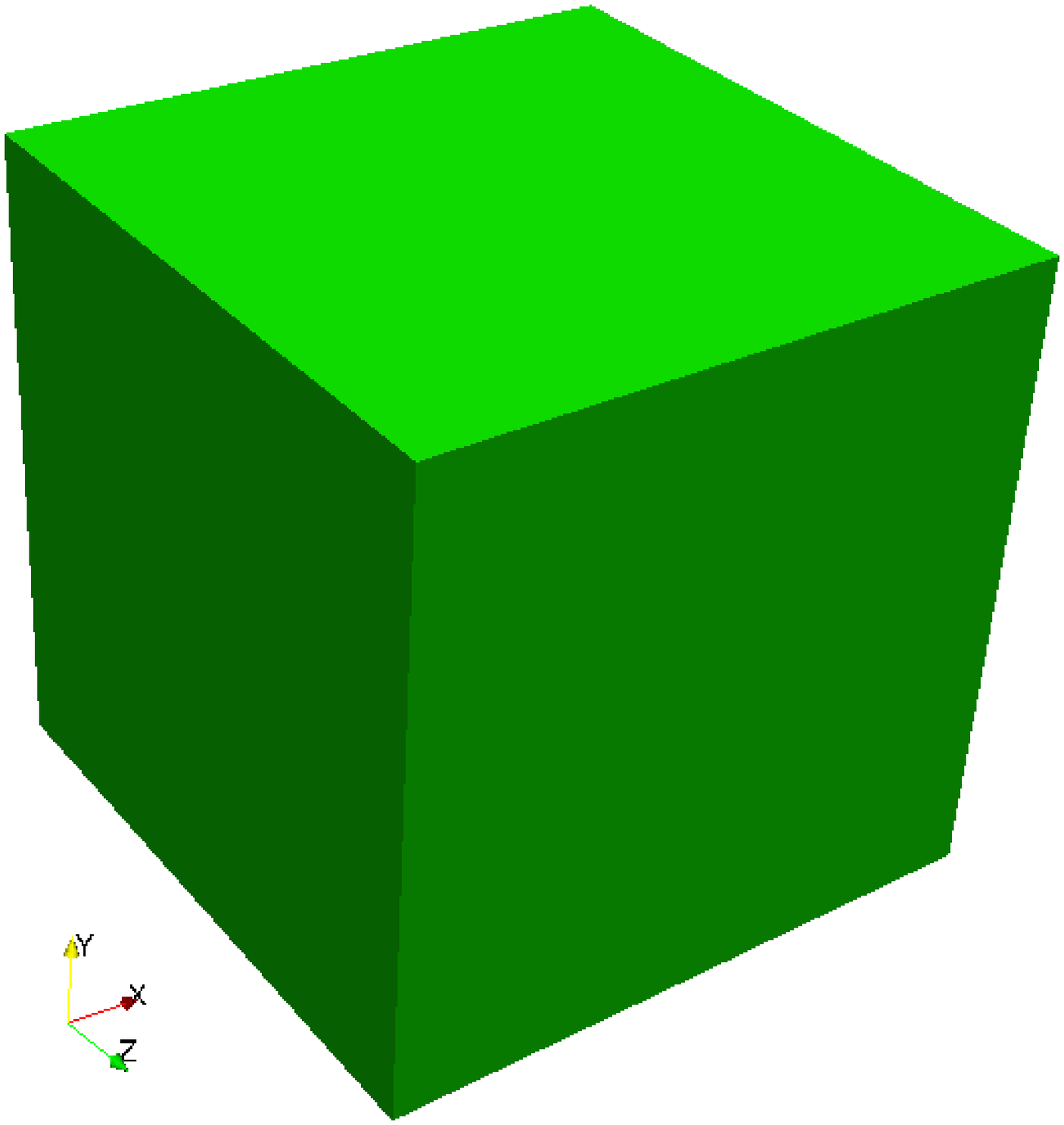} 
	   \label{fig:loadvy2}
	}	
	\subfigure[M$_3$]
	{
	\includegraphics[width=0.25\textwidth]{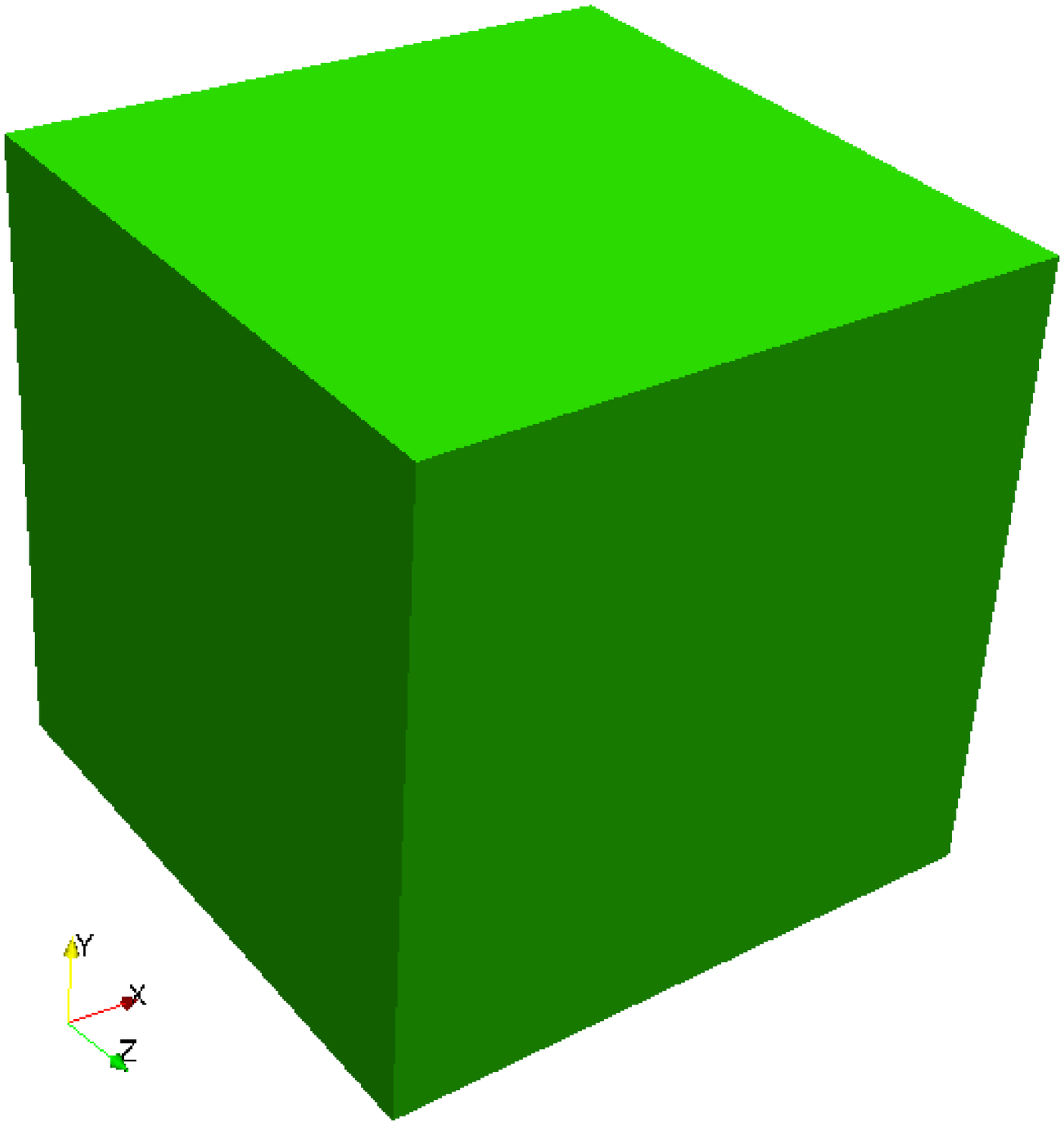} 
	   \label{fig:loadwz2}
	}	
	\subfigure
	{
	\includegraphics[width=0.25\textwidth]{spectrum.eps} 
	}			
	\caption{(Color online) Microstructure evolution during RT:  the first row show the microstructure at $ \bar{t} = 950 $ and the second row at the end of loading cycle  $ \bar{t} = 1080 $ (red color represent  M$_i $ variant, blue represent the remaining two variants M$ _j $ and M$_k $, and green color represents austenite (A) phase).}
    \label{fig:loading}    
\end{figure}

Next, we perform a temperature induced reverse transformation (RT, martensite $ \rightarrow $ austenite) starting with the evolved microstructure in the previous step. The thermal loading is applied on a domain with $ \bar{g} = 0.05 \bar{t}  $ in the dimensionless units (bar shows the dimensionless variable). Fig. \ref{fig:loading} shows the time snapshot of the microstructure at intermediate time (first row) and at the end of unloading (second row). The domain walls no longer remain distinct and sharp, as compared to Fig. \ref{fig:MicrostructureEvolve}, and extinct at the end of thermal loading. This observation is in accordance with the experimental evidence \cite{Lagoudas}.

\begin{figure}[H]
	\centering
	\subfigure[M$_1$]
	{
	\includegraphics[width=0.25\textwidth]{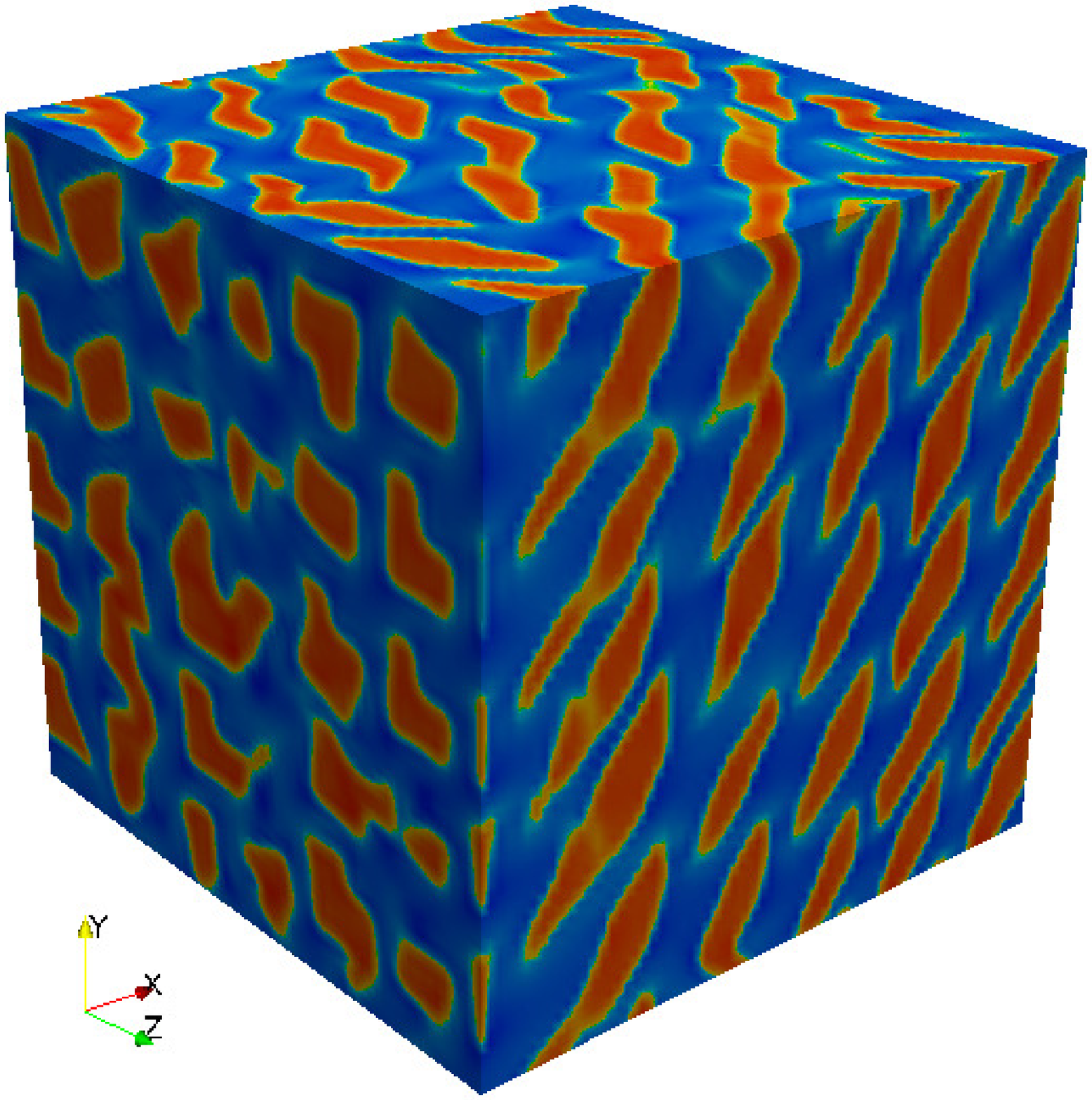} 
	   \label{fig:unloadux}
	}
	\subfigure[M$_2$]
	{
	\includegraphics[width=0.25\textwidth]{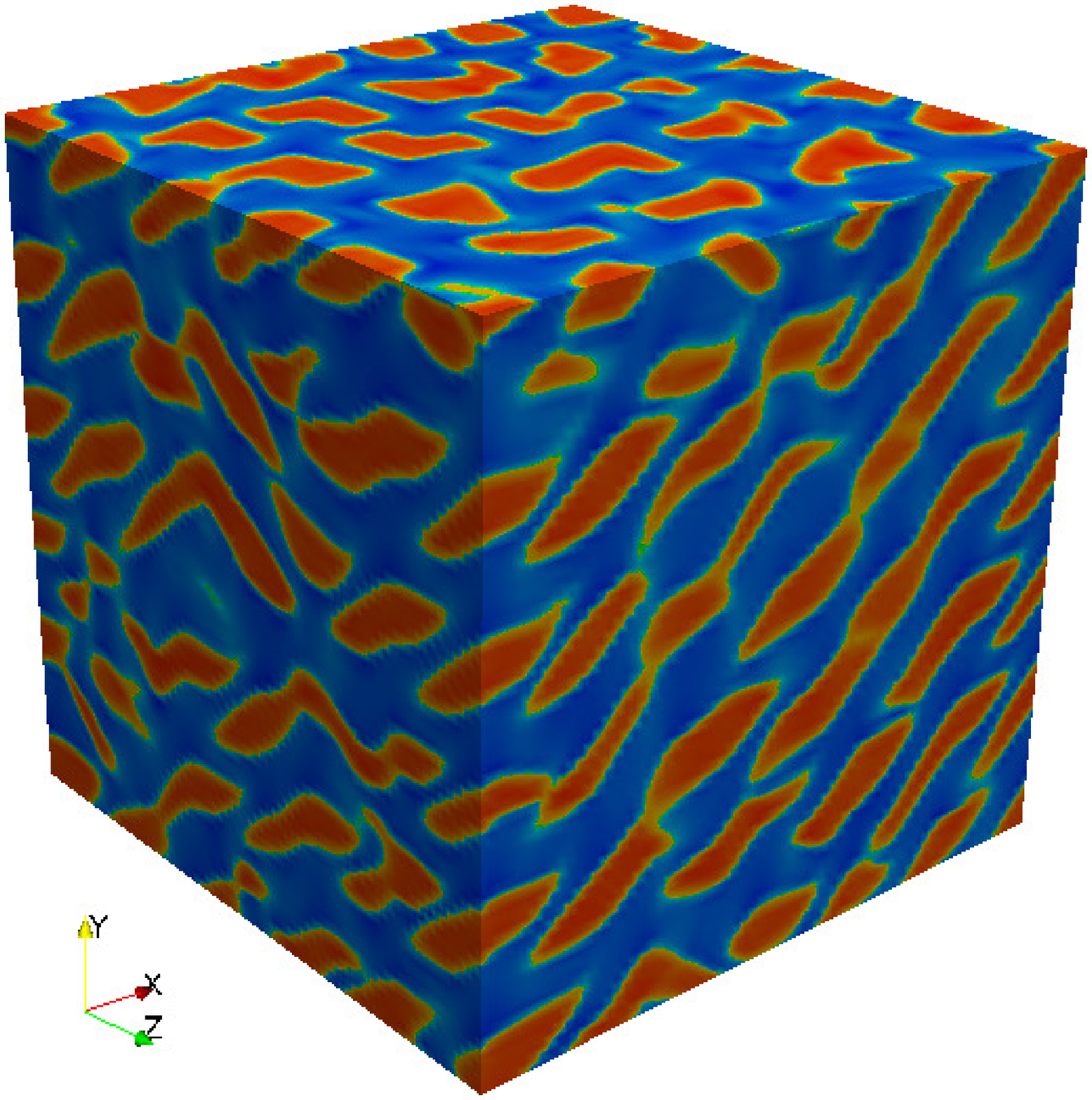} 
	   \label{fig:unloadvy}
	}	
	\subfigure[M$_3$]
	{
	\includegraphics[width=0.25\textwidth]{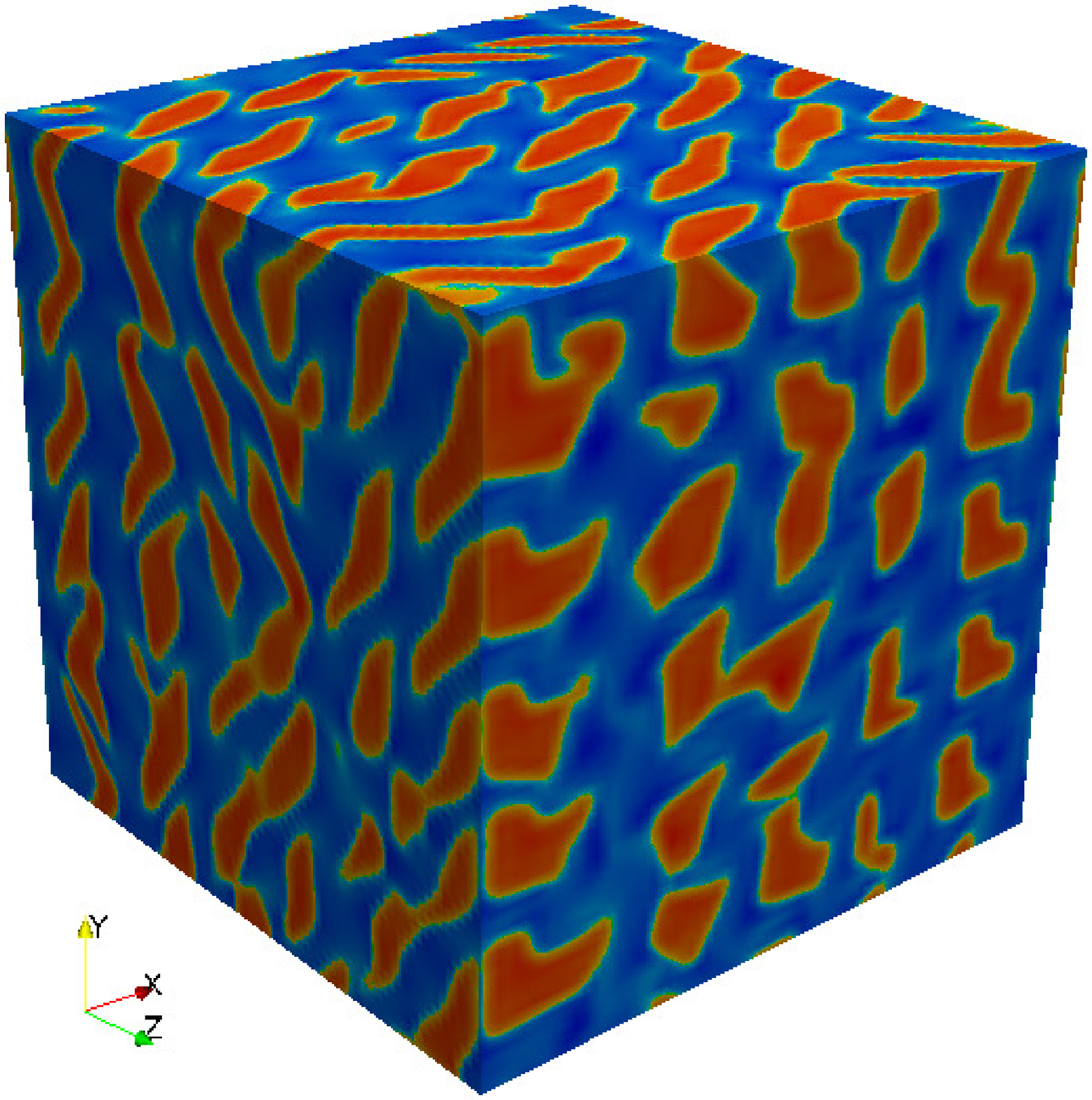} 
	   \label{fig:unloadwz}
	}	
	\subfigure
	{
	\includegraphics[width=0.25\textwidth]{spectrum.eps} 
	}			
	\caption{(Color online) Transient microstructure at $ \bar{t} \approx 1500 $ during the FT (red color represents  M$_i $ variant, blue represents the remaining two variants M$ _j $ and M$_k $, and green color represents austenite (A) phase).}
    \label{fig:unloading}    
\end{figure}

Finally, we use the evolved austenite microstructure at the end of loading cycle as the initial condition to the forward transformation (FT, austenite $ \rightarrow $ martensite) by applying the thermal loading on a domain with $ \bar{g} = -0.1 \bar{t}  $ in the dimensionless units. The martensitic microstructure evolve abruptly at $ \tau \approx -5 $ at approximately 1500 time units during unloading. The transient martensitic variants on nucleation are shown in Fig. \ref{fig:unloading}. 

\begin{figure}
	\centering
	\includegraphics[width=0.5\textwidth]{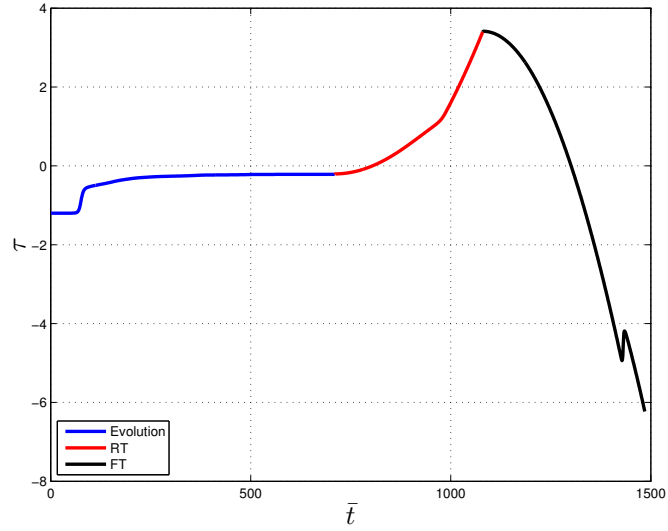}
	\caption{(Color online) Average temperature coefficient $ \tau $ plot during microstructure evolution (blue), RT (red), and FT (black).}
    \label{fig:TimeVsTemp} 	
\end{figure}

The average temperature coefficient $ \tau $ evolution in SMA domain during microstructure evolution, RT, and FT are shown in Fig. \ref{fig:TimeVsTemp}. The nucleation of martensitic variant from austenite during FT is seen with a jump in $ \tau $ at approximately 1500 time units. 

\section{Conclusions} \label{sec:Conclusions}

The fully coupled thermo-mechanical model to describe cubic-to-tetragonal phase transformations in SMAs has been developed and numerically implemented in the isogeometric analysis. 
We have numerically analyzed the temperature induced reverse and forward phase transformations in SMAs. It has been found that the domain walls between martensite variants are aligned in accordance with the crystallographic theory and experimental results. We have also captured the abrupt nucleation of martensitic variants during the reverse transformation.

\section*{Acknowledgement}
This work was made possible with the facilities of the Shared Hierarchical Academic Research Computing Network SHARCNET: www.sharcnet.ca) and Compute/Calcul Canada. Support of NSERC and CRC program is also gratefully acknowledged.

\begingroup
\setstretch{0.8}
\bibliographystyle{unsrt}
\bibliography{AMMCS}
\endgroup

\end{document}